\documentclass[preprint,12pt]{elsarticle}

\usepackage{amssymb}

\journal{Nuclear Instruments and Methods A}

\begin{document}

\begin{frontmatter}



\title{Data encoding efficiency in pixel detector readout with charge information}


\author[lbl]{Maurice Garcia-Sciveres}
\author[uch]{Xinkang Wang}
\address[uch]{University of Chicago, Chicago, IL, USA.}
\address[lbl]{Lawrence Berkeley National Laboratory, Berkeley, CA, USA.}

\begin{abstract}
The average minimum number of bits needed for lossless readout of a pixel detector is calculated, in the regime
of interest for particle physics where only a small fraction of pixels have a non-zero value per frame. 
This permits a systematic comparison of the readout efficiency of different encoding implementations. 
The calculation is compared to the number of bits used by the FE-I4 pixel readout chip of the ATLAS experiment.  
\end{abstract}


\end{frontmatter}

\section{Introduction}

In reference~\cite{stripreadout} we presented a prescription for calculating the efficiency of readout encoding methods for binary strip detector readout, which is arguably the simplest case of interest for particle physics instrumentation. Here we extend the analysis to pixel detectors and also to include charge information rather than binary readout. 
Both the two-dimensional nature of pixel hit patterns and the addition of charge information introduce significant complexity. 
We preserve the same meaning of readout efficiency as the number of bits used in practice to extract all the required information from the detector, 
relative to the minimum possible number of bits needed given by the information content. Calculating this 
minimum is the main content of this note. 
This does not mean the bits used for a single 
event or readout cycle, but the average bits per event for a large ensemble of events.
In practice we define efficiency as the minimum possible number over the actual number of bits, 
so that it has a value between 0 and 1. We consider only lossless data compression, as opposed to data reduction in which information is discarded and cannot be recovered. 

We consider typical pixel detector occupancy below 0.5\%, where occupancy always means average occupancy over a fairly long time (eg. 1 second). The average occupancy is what determines the output data total readout bandwidth requirement, rather than the instantaneous occupancy of a single event. The actual readout bandwidth required for a given occupancy depends on how much the data can be compressed, and also on the required latency (how long one is willing to wait for the information). We only consider the compression aspect here; for a case study that includes latency considerations see ref.~\cite{wit2014}. 
An important practical constraint is that the detector readout will be implemented in units (such as modules or chips). 
The information content of the detector is the sum of the information in all the readout units. We are therefore 
calculating the entropy for one single readout unit. If the data from the entire detector could somehow be combined prior 
to readout, then the entropy may be lower than the plain sum of all readout units. But we restrict our analysis to 
assume independent readout units as the case of practical interest.

\begin{figure}[ht]
\centerline{\includegraphics[width=0.7\textwidth]{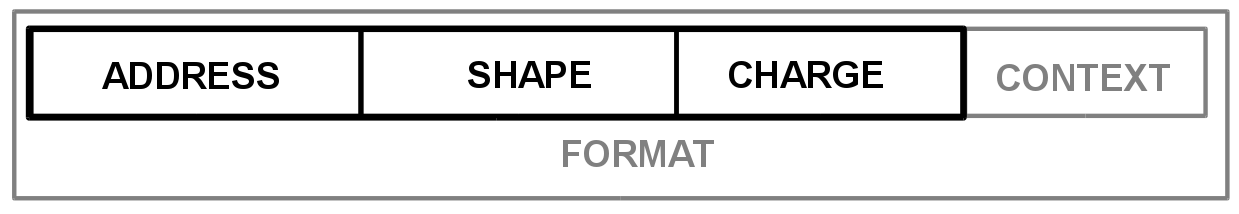}}
\caption{Schematic breakdown of the detector output data into component parts. The three bold face components 
are analyzed in this note.}
\label{fig:parts}
\end{figure}

Following ref.~\cite{stripreadout} we decompose the detector output data into the parts shown in Fig.~\ref{fig:parts}, 
where we have added cluster shape and charge elements that were not present for binary strip detector readout. 
A cluster is any combination of hit pixels which are {\em touching}, including corner-to-corner touching. Any two 
hit pixels that touch belong to the same cluster. 
We will only study the bold face parts in this note, since the treatment of context and format elements given in 
ref.~\cite{stripreadout} remains valid. 
We have chosen this decomposition for convenience, but other decompositions are possible. 
The desired end result is the minimum number of bits necessary to encode all the information, or information entropy,
which is a physical property and should not depend on the representation chosen to calculate it. 
However, correlations in real data that are not included in our assumptions will introduce a bias and 
result in a representation-dependent entropy. This is discussed further in Section~\ref{sec:assumptions}. 

The decomposition of Fig.~\ref{fig:parts} has a first order physical interpretation as follows.
The {\em addresses} correspond to the positions of particles crossing a silicon sensor- 
each particle forming one cluster that is identified by one address
(the cluster address can be defined in several ways, for example as the bottom left corner pixel of the 
smallest cluster bounding box. How it is defined is not important to our discussion). 
If the distribution of particle tracks crossing a sensor is uniform,
the cluster address entropy is straightforward to calculate. 
Each cluster has a {\em shape} and {\em size}, which are given by
cluster distribution functions which depend on position, orientation, and other features that may or may not vary across
a particular detector. 
There are nevertheless a finite number of 
possible shapes and sizes. Finally, there is a {\em charge} distribution in each cluster. In the classical limit charge is an analog quantity and therefore a potential problem for an entropy calculation, which requires a discrete distribution.  
However, the practical measurement of charge is limited by
noise, and we must therefore consider the entropy of encoding useful charge information, not meaningless noise fluctuations. 
For this we will further decompose charge into total charge, $QT$, and single pixel charge fractions, $QF$. 
To higher order, multiple
particles can merge into one cluster, secondary interactions from a single particle can result in multiple clusters, 
and dead pixels or charge deposits that fluctuate below threshold can lead to split clusters. 
So the association of clusters to randomly incident
single particles is not perfect, but as we will see at low occupancy such high order corrections have a negligible impact
on the entropy. 

Sections \ref{sec:address}, \ref{sec:shape}, and \ref{sec:charge} evaluate and discuss the address, shape, and charge entropy, respectively. The total information entropy due to the hits is the sum of these parts, 
\begin{equation}
\label{hhits}
H_{\rm hits} = H_A + H_s + (H_{QT} + H_{QF})
\end{equation}
which excludes the context and format contributions as already explained. 

\section{Assumptions}
\label{sec:assumptions}

This paper calculates an information entropy by making general assumptions about the data 
from particle physics pixel detectors. For the result to be applicable to 
a particular pixel detector, one must first check that the assumptions are valid
for the detector in question. This section discusses how the assumptions may fail and what 
would be the consequence on the result. This section refers to material in the 
sections that follow, but we placed this discussion first, so the 
reader is aware of the issues and can refer back to this section as needed. 

The address entropy calculation (Sec.~\ref{sec:address}) 
assumes clusters are mainly due to particles that illuminate 
a readout unit uniformly and randomly. This may not be the case, for example in  
collimated particle jets, where the clusters themselves may be ``clustered'' (spatially correlated)
on the scale of a readout unit. Correlations between cluster positions would reduce the address 
entropy. For the particular case of a high luminosity proton collider with many interactions 
per beam bunch crossing (pile-up), the vast majority of clusters will come from pile-up interactions, 
with high energy collimated jets being rare. Therefore, we expect that the uniform, random assumption
will hold. Note that what matters for entropy is the vast majority, not the rare exceptions. 

Sec.~\ref{sec:address} also assumes that most clusters are produced by one single particle, 
with merging of energy deposits
from two or more particles being rare. This of course depends on the track density and the detector design.
A low granularity detector in high track density, or used for collimated beams or jets (eg. a calorimeter) 
may have most clusters encompass multiple particles. Thus our assumption is appropriate for tracking 
detectors, which, in order to be useful, must be designed with high enough granularity for the 
assumption to hold. 

The separate calculations of Sections \ref{sec:address}, \ref{sec:shape}, and \ref{sec:charge}
assume that the entropy can be decomposed exactly into the parts shown in Fig.~\ref{fig:parts}.
That is, we have not considered any correlation between cluster positions, cluster charge, pixel charge fraction,
and cluster shape. But such correlations do exist at some level and will result in a reduction of 
entropy. Consider a flat (not curved) detector element illuminated by a point source of particles. 
Each cluster position in the detector will be associated with a different particle incidence angle, 
and therefore a different cluster shape and charge. We have ignored such correlations within
a readout unit, assuming that a readout unit is small, so that even for a point source there would be little 
variation. Furthermore, if the beam spot is not a point source, but in fact physically larger than the 
readout unit, any correlation between position and incidence angle will be washed out (Recall that 
our calculation is for one readout unit, so only correlations within a readout unit matter.)

In Sec.~\ref{sec:charge} we have ignored correlation between cluster charge and pixel charge fractions. 
When the charge of an n-pixel cluster takes on it smallest (biggest) possible value, then the pixel charge fractions 
are given: they must all be equal and given by the pixel threshold (maximum). 
As cluster charge increases (decreases), more 
and more pixel fraction combinations become possible, and correlation disappears. 
This type of correlation lowers the entropy. 

The above discussion concerns assumptions that omit correlations thought to 
be small for tracking detectors in high intensity colliders. As discussed, the effect of these 
correlations will be to lower the entropy. Therefore, to the extent that the assumptions are violated,
the minimum number of bits calculated in this paper can be seen as an upper limit to the true minimum number of bits.
Other assumptions may instead increase the entropy if violated.
Secs.~\ref{sec:shape} and ~\ref{sec:charge} assume a detector is exposed to a single dominant source of particles 
(originating form a single interaction region and containing a mix of particles with a peaked ionization distribution, 
such as for minimum ionizing particles). 
If a detector is instead exposed to multiple, comparable intensity sources at the same time, 
that would complicate the estimate and result in higher entropy (weak secondary sources will not matter,
as the entropy is determined by the bulk of the data, not by rare events).

\section{Address Entropy}
\label{sec:address}

\begin{figure}[ht]
\centerline{\includegraphics[width=0.7\textwidth]{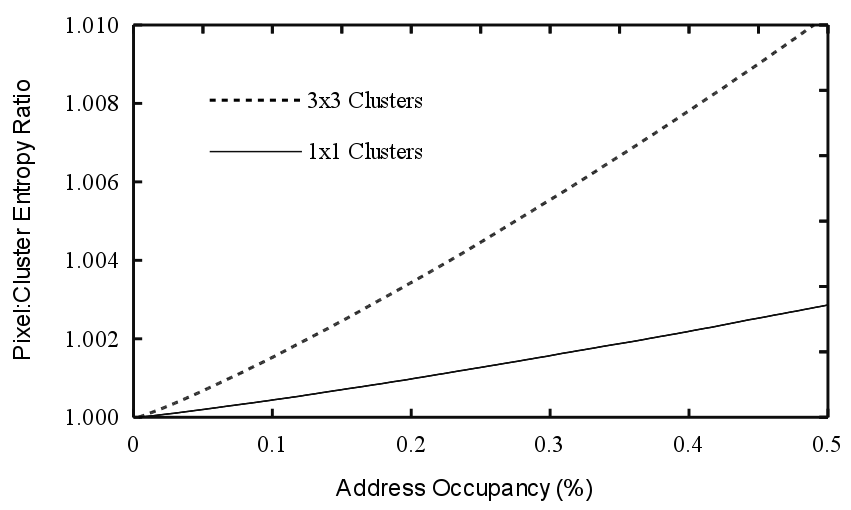}}
\caption{Ratio of single pixel address entropy to entropy of different size isolated clusters, as explained in the text. 
This figure is for a total number of pixels $n=2^{15}$.}
\label{fig:HAratio}
\end{figure}

In the limiting case of individual pixels rather than clusters, 
the address entropy is given simply by the logarithm of the number of ways to pick 
$j$ out of $n$ pixels, where $j/n$ is the occupancy and $n$ is the number of pixels in the chip. This is given by the expression $log_{_2}$$n \choose j$. When clusters are considered, two clusters cannot be touching, or they would count as a single cluster. One must therefore exclude a 1-pixel-wide empty boundary around each cluster.
Let $z$ be the total number of pixels ``used up'' by a cluster {\em including} this empty boundary.
For example, for a 1-pixel cluster, $z=9$ because the boundary around 1 isolated pixel consists of 8 pixels. 
For a 3 by 3 pixel cluster, $z=25$, and so on. 
The address entropy then has a lower bound
\begin{equation}
\label{eq:Haddress}
H_z = log_{_2} [ \Pi^{k-1}_{i=0}(n-zi)/(i+1) ]
\end{equation}
where $k$ is the number of clusters. 
This of course reduces to $log_{_2}$$n \choose k$ for $z=1$ (1-pixel clusters with no empty boundary). 
$H_z$ is a lower bound because 
at the edges of a chip, or when two clusters are very close, a dedicated empty boundary for each cluster is not needed. 
Fig.~\ref{fig:HAratio} shows the ratio 
$log_{_2}$$n \choose k$$/H_z$ for $z=9$ (1-pixel clusters plus an empty boundary) and 
$z=25$ (3 by 3-pixel clusters plus an empty boundary). It can be seen that, for the occupancy range of interest, 
$log_{_2}$$n \choose k$ and $H_z$ are very close, and therefore $H_A=log_{_2}$$n \choose k$  is a 
good approximation to the true entropy. We will use $H_A$ as the address entropy in our calculations.  

\begin{figure}[ht]
\centerline{\includegraphics[width=0.7\textwidth]{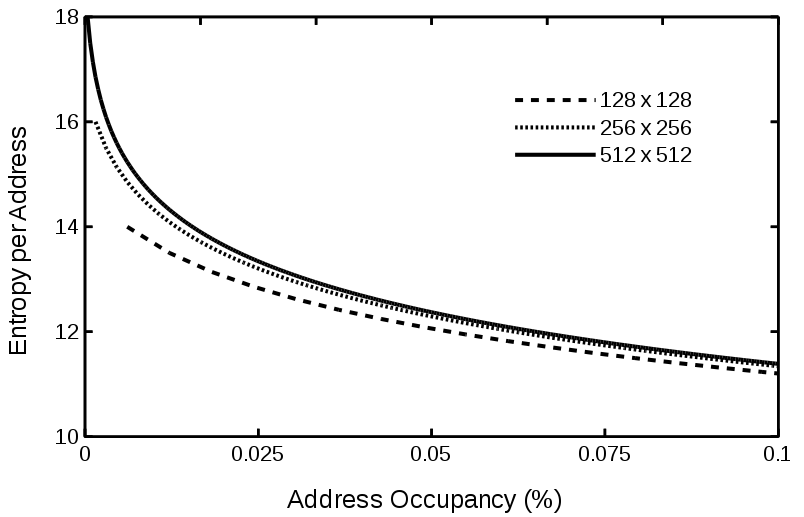}}
\caption{Address entropy per address as a function of address occupancy, for different size chips as indicated.}
\label{fig:HAperA}
\end{figure}
  
It is instructive to examine the address entropy per address as a function of occupancy. For a single address, the entropy is simply the number of bits needed to count the full address space, so for a readout chip with $2^{16}$ pixels the address entropy of a single address (occupancy of $2^{-16}$) is 16 bits. But as the occupancy rises, fewer and fewer bits are needed per address. Within the occupancy range of interest, the number of bits per address quickly converges to a common value 
independent of chip size. Fig.~\ref{fig:HAperA} shows the address entropy per address as a function of occupancy for chips sizes $2^{14}$, $2^{16}$, and $2^{18}$ pixels 
(note that the occupancy range in this figure goes only up to 0.1\% so that the difference between the curves can be appreciated). This is an important result counter to conventional wisdom. 
It shows that making a chip larger and therefore growing the address space (keeping pixel size constant) 
does {\em not} imply a greater data volume due to the need for more address bits. 
If addresses are compressed (rather than using the same number of bits per address regardless of occupancy), 
then approximately the same number of bits is needed to transmit the cluster address information regardless of chip size. 

Making pixels smaller, on the other hand, does 
increase the entropy per address as expected 
(i.e. cutting the pixel size in half adds one bit). This is more difficult to appreciate from Fig.~\ref{fig:HAperA}, 
where reducing pixel size has the effect of reducing occupancy (so moving to the left on the curves) 
as well as increasing the number of pixels for constant size chip.

\section{Shape Entropy}
\label{sec:shape}

\begin{figure}[h]
\centerline{\includegraphics[width=0.7\textwidth]{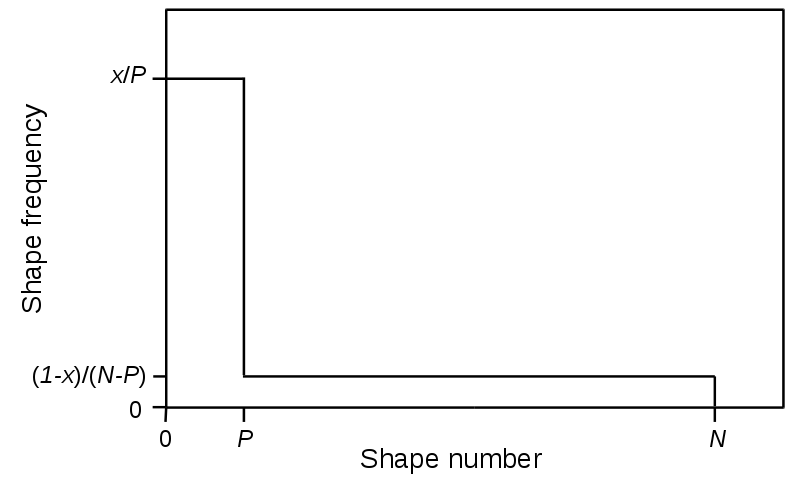}}
\caption{Representation of distribution of cluster shapes in an arbitrary readout unit}
\label{fig:shapes}
\end{figure}
\begin{figure}[h]
\centerline{\includegraphics[width=0.7\textwidth]{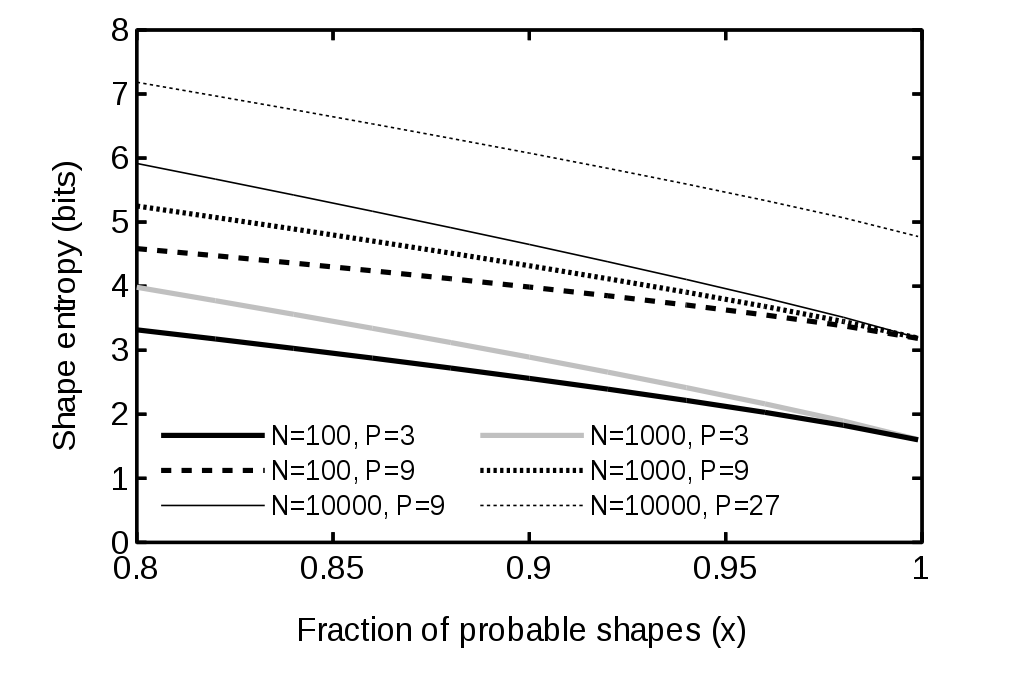}}
\caption{Shape entropy for different choices of $N$, the total number of cluster shapes, and $P$, 
the number of most frequent shapes making up the total fraction shown on the $x$ axis.}
\label{fig:shape-entropy}
\end{figure}

We consider shape to also include the size (number of pixels) of a cluster. A single pixel has zero shape entropy, 
but two adjacent pixels can have 4 possible shapes (up-down, left-right, and two diagonals). 
Three adjacent pixels can have 20 possible shapes, four pixels 110 shapes, 
and so on (\cite{polykings} discuses the counting of cluster shapes for Manhattan geometry). 
From this it seems that shape entropy can be rather large, and difficult to estimate in general. 
But in real-life particle detectors this will not be the case, and we can estimate a shape entropy without knowing a detailed cluster zoology. 
The reason is that each readout element of a pixel detector (a single chip or a multi-chip module) 
will be traversed by particles from a preferential direction, and these will produce similar-looking clusters. 
Thus, we can consider that some large fraction, $x$, of the clusters in a given readout unit consist of a small number of shapes, 
$P$, while the remaining $1-x$ can have a large variety of shapes, $N-P$, due to scattered particles and radiation. 
This is shown graphically in Fig.~\ref{fig:shapes}.
The shape entropy is thus,
\begin{equation}
H_s = -xlog_2(x/P) - (1-x)log_2(\frac{1-x}{N-P})
\end{equation}
$H_s$ is plotted for different $N$ and $P$ in Fig.~\ref{fig:shape-entropy}. Not surprisingly, the shape entropy is dominated
by the number of cluster shapes that occur very frequently, and the presence of even thousands of additional, but improbable 
shapes has little impact. For estimation purposes we will take 4 bits per cluster as a reasonable indicative value for 
shape entropy in present day pixel detectors where the mean cluster size is small. However, the cluster shapes occurring frequently in one particular module of a pixel detector will be different from those in another module with different position relative to the collisions. 
Therefore, in order to realize in practice low entropy encoding of cluster shapes, each module would need to use a different encoding (a different Huffman table, for example), which would have to be programmed or learned by the module from its own data.

\section{Charge Entropy}
\label{sec:charge}

We first consider cluster charge. 
The ideal cluster charge obeys a Landau distribution for a minimum ionizing particle passing through a silicon detector. 
One must consider a spectrum of different particle momenta and incidence angles, 
which leads to a sum of Landaus, as well a Gaussian broadening due to noise. 
As was the case for cluster shapes, the range of Landau means to be summed will vary for modules 
in different detector locations, but each readout unit will have 
a known, Landau-like cluster charge distribution.
To estimate the entropy we consider a cluster charge probability distribution function $QT$ equal to the
sum of 5 Landau distributions with means at $\mu =$4, 5, 6, 7, and 8 arbitrary 
units, where the Landau function is $QT(q) = \frac{1}{\sqrt{2\pi}}e^{-[q-\mu+e^{-(q-\mu)}]/2}$.
Using a single Landau instead, or changing the mean, made very little difference. 
The mean of $QT$ is just the average of the component means, or 6.5.  

\begin{figure}[ht]
\centerline{\includegraphics[width=0.7\textwidth]{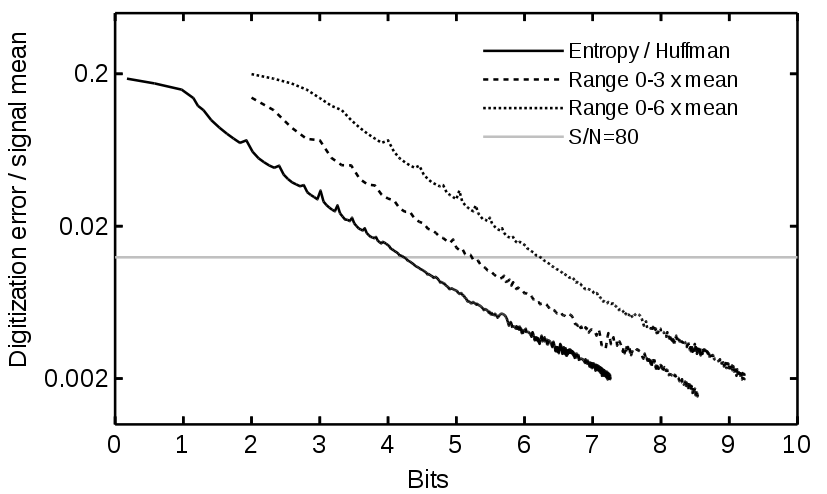}}
\caption{Digitization error for average of 4 measurements (4 layers) vs. the entropy value (in bits) or 
average number of bits per cluster. 
The 0-3 x mean and 0-6 x mean ranges refer to the digitization range for which the binary ADC value is transmitted without compression. 
Compressing the ADC value using Huffman codes achieves the entropy value regardless of the ADC range. }
\label{fig:clusterq}
\end{figure}

To compute an entropy we first ``digitize'' the charge PDF by histogramming it in $2^D$ equal width bins in the range 
0 to 20 (40) arbitrary units. The last (overflow) bin is increased so that the total histogram probability is unity. 
Note that 20 (40) corresponds to just over 3 (6) times the mean of this particular PDF, meant to
be representative of present pixel detectors with a dynamic range of a few times the minimum ionizing particle signal~\cite{ibl,cmsphase1}. 
For each value of $D$ we compute the entropy as $H = -\sum_i p_i log_{_2}(p_i)$, where the sum is over all histogram bins
and $p_i$ is the probability in each bin. We also compute the {\em digitization error} by performing a toy experiment 
in which we measure cluster charge in 4 hypothetical layers and take the average. This is representative 
of measuring the specific ionization of particle tracks in a silicon detector. The true value for each layer is randomly
drawn from $QT$, 
while the measurement is taken to be the central value of the histogram bin that the true value falls into. 
The digitization error is the standard deviation of the difference between the average of the 4 measurements and the true average. The digitization error vs. cluster charge entropy, $H_{QT}$, is shown in Fig.~\ref{fig:clusterq}. 

As previously mentioned, we want to calculate the entropy for a meaningful charge measurement, 
and the most precise possible meaningful measurement is limited by noise. 
A horizontal line in Fig.~\ref{fig:clusterq} is included to represent the noise level. 
Obviously the noise level will vary from system to system, so this value is an example. We drew the line at 0.0125
which means S/N=80 for the average of 4 measurements (4 layers with cluster S/N=40 per layer).  
This shows that for this particular noise level the cluster charge entropy 
is $H_{QT}=4.2$ bits per cluster. Fig.~\ref{fig:clusterq} also includes two additional curves to show the 
bits per cluster for an uncompressed linear ADC scale for the two dynamic ranges considered: 20
units and 40 units (the dynamic range has no effect on the calculated entropy). We also explicitly compressed the 
ADC values using Huffman codes, and the average number of bits after compression differed from 
the calculated entropy by at most 5\% for both dynamic ranges (would not able to see as a separate line if plotted 
on the figure).

In addition to cluster charge, we must also consider the charge of individual pixels in a cluster. 
The single pixel charge distribution does not have a universal Landau form. As we already have 
analyzed the cluster charge, the remaining information is not the absolute charge, 
but the fraction of the total cluster charge in each pixel. The analysis is particularly simple for
a 2-pixel cluster. As both pixels must have the same charge fraction distribution $F$, it follows 
that $F$ must be symmetric $F(r)=F(1-r)$, where the fraction $r$ ranges between 0 and 1. 
Furthermore, in the ideal case that the charge splitting is due to the 
ionizing particle that created the cluster crossing the boundary between the pixels
(as opposed to a systematic effect like electronic crosstalk), then there is no favored splitting
and the ratio distribution is simply $F=1$. Thus, F is digitized in $2^D$ equal size bins with equal probability in
each bin and the entropy is simply $D$/2 per pixel 
($D$ is divided by 2 because only the charge fraction in one pixel need be specified).
The smallest useful bin size should be simply given by the single pixel noise over the single pixel average 
signal. Thus, for 2-pixel cluster S/N=40, the signal to noise on the single pixel fraction will be 
approximately 20. Hence, $H_{QF}$ = $log_2(20)/2=2.15$ per pixel. We will take this as a general estimate 
on average. Clearly for individual clusters $H_{QF}$ will vary: $H_{QF}=0$ for 1-pixel clusters, and
$H_{QF}>2 \times 2.15$ for more than 2-pixel clusters.

\section{Combined Result}
We can now combine the above estimates to obtain a value for $H_{\rm hits}$ of Eq.~\ref{hhits}, which 
we write as an entropy per cluster.
From Fig.~\ref{fig:HAperA} we see that $H_A/N_{\rm cluster} \approx 12$. 
From Fig.~\ref{fig:shape-entropy} we can estimate that $H_s/N_{\rm cluster} \approx 4$. 
From Fig.~\ref{fig:clusterq} $H_{QT}/N_{\rm cluster} = 4.2$. 
Also $H_{QF}/N_{\rm pixel}=2.15$. All these estimates are independent of fine details about the 
detector in question other than signal/noise.  
One parameter needed is the average number of hit pixels per cluster, due to $H_{QF}$. Taking 
a typical value of $N_{\rm pixel}/N_{\rm cluster} \approx 2$ yields,
\begin{equation}
H_{\rm hits}/N_{\rm cluster} \approx 24.5
\end{equation}

As a concrete example we can compare this to the number of bits 
per cluster used by the FE-I4 chip of the ATLAS experiment~\cite{fei4}, excluding format and context information, 
for an average cluster size of 2. 
For a $2^{16}$ ($2^{18}$) pixel chip, the number of bits for a 1-pixel cluster 
used by the FE-I4 encoding is 24 (26). For a 2-pixel cluster it can be 24 (26) or 48 (52), depending on the 
shape, and for a 3-pixel cluster 48 (52) or 72 (78).
The FE-I4 encoding includes 4 bits of uncompressed ADC value per pixel, which is not
sufficient for a S/N=40 cluster charge measurement even with limited dynamic range (Fig.~\ref{fig:clusterq}).  
The average number of bits depends on the cluster distributions. Using ATLAS experiment inner layer 
distributions~\cite{clusters} yields an average of 35 (37) bits per cluster.  

\section{Conclusion and Outlook}

We have shown that it is possible to estimate the entropy of pixel detector hit data in the occupancy range 
of interest to particle physics, without knowing all the details about a specific detector. Approximate 
knowledge about the occupancy, signal to noise, and cluster size distributions is sufficient.
This is useful to understand how much room for improvement there is when developing a new 
detector readout. In particular, new pixel detectors for the High Luminosity LHC~\cite{atlasphase2A,atlasphase2B,cmsphase2}
will have very high data volume, requiring efficient encoding of the information to be transmitted.
A comparison to the readout encoding used by the ATLAS FE-I4 shows that, even with reduced signal to noise
capability, the FE-I4 encoding has of order 40\% total readout bandwidth to be gained from data compression. With smaller pixels 
and therefore broader cluster size distributions, the gains from applying on-chip data compression
will likely be even greater in future detectors.  

\section*{Acknowledgements}
This work was supported in part by the Office of High Energy Physics of the U.S. 
Department of Energy under contract DE-AC02-05CH11231.





\end{document}